\documentclass[twocolumn]{aa}
\usepackage{graphicx}
\usepackage{natbib}
\usepackage{txfonts}
\begin{document}

\title{Geometrical tests of cosmological models}
\subtitle{I. Probing dark energy using the kinematics of high-redshift galaxies}

\author{
C. Marinoni$^{1}$, A. Saintonge$^{2}$,
R. Giovanelli$^{2}$,  M.P. Haynes$^{2}$,  K.L. Masters$^{3}$, 
O. Le F\`evre$^4$, \\  A. Mazure$^{4}$, P. Taxil$^{1}$, J.-M. Virey$^{1}$}

\offprints{C. Marinoni, \email{marinoni@cpt.univ-mrs.fr}}

\institute{
$^1$ Centre de Physique Th\'eorique\thanks{Centre de Physique Th\'eorique is UMR 6207 -
 ``Unit\'e Mixte de Recherche'' of CNRS and of the Universities ``de Provence'',
 ``de la M\'editerran\'ee'' and ``du Sud Toulon-Var''- Laboratory
 affiliated to FRUMAM (FR 2291).}, CNRS-Universit\'e de Provence, Case 907,
 F-13288 Marseille, France.\\
$^2$ Department of Astronomy, Cornell University, Ithaca, NY 14853, USA\\
$^3$ Harvard-Smithsonian Center for Astrophysics, Cambridge, MA 02143, 
USA\\
$^4$ Laboratoire d'Astrophysique de Marseille, UMR 6110, CNRS Universit\'e 
de Provence, 13376 Marseille, France \\
}

\date{Received 17 January 2007 / Accepted 2 August 2007}

\authorrunning{Marinoni et al.}
\titlerunning{Geometrical tests of cosmological models. I. }

\abstract{

We suggest  to  use the   observationally measured  and  theoretically
justified correlation between size and rotational velocity of galactic
discs as a  viable method to select a   set of high redshift  standard
rods which may  be  used to explore the   dark energy  content of  the
universe via the classical angular-diameter  test.  Here we explore  a
new strategy for  an optimal implementation  of this test.  We propose
to use the rotation speed of high redshift galaxies as a standard size
indicator and  show how high  resolution multi-object spectroscopy and
ACS/HST high  quality spatial images,  may be combined  to measure the
amplitude of the  dark  energy density  parameter  $\Omega_{Q}$, or to
constrain  the cosmic equation  of state  parameter  for a smooth dark
energy   component  ($w=p/\rho, \;\; -1\le  w  <  -1/3$).  Nearly 1300
standard rods with high velocity rotation in the bin $V=200\pm 20$km/s
are expected in a field of 1 sq. degree and over the redshift baseline
$0<z<1.4$. This sample is sufficient  to constrain the cosmic equation
of state parameter  $w$ at a level   of $20\%$ (without  priors in the
[$\Omega_m,\Omega_Q$]  plane)  even  when  the [OII]$\lambda 3727$\AA~
linewidth-diameter relationship is  calibrated with a scatter of $\sim
40\%$. We evaluate how systematics may  affect the proposed tests, and
find that a  linear standard rod  evolution, causing galaxy dimensions
to be up to $30\%$ smaller at $z=1.5$,  can be uniquely diagnosed, and
will   minimally   bias  the  confidence     level contours   in   the
[$\Omega_{Q}$, $w$] plane.  Finally, we   show how to derive,  without
{\it a priori} knowing the specific functional form of disc evolution,
a cosmology-evolution diagram with which it is possible to establish a
mapping between different cosmological models and the amount of galaxy
disc/luminosity evolution expected at a given redshift.

 }

\maketitle

\section{Introduction}

Several  and    remarkable progresses  in  the  understanding   of the
dynamical status of the universe,  encourage us to believe that, after
roaming from paradigm to paradigm, we are finally converging towards a
well-founded, internally consistent standard model of the universe.

The  picture emerging from  independent  observations and  analysis is
sufficiently coherent  to   be referred to  as   the {\it concordance}
model \citep[e.g.][]{teg06}. Within this  framework,  
the universe is  flat  ($\Omega_K = -0.003^{+0.0095}_{-0.0102}$)  
composed   of  $\sim 1/5$ cold dark matter ($\Omega_{cdm} \sim 0.197^{+0.016}_{-0.015}$) and   
$\sim 3/4$ dark energy  ($\Omega_{\Lambda} = 0.761^{+0.017}_{-0.018}$),  with large
negative pressure ($w =-0.941^{+0.017}_{-0.018}$), and with a very low baryon content 
($\Omega_b = 0.0416^{+0.0019}_{-0.0018}$).  
Mounting and compelling evidence for accelerated expansion of
the universe, driven by a  dark energy component, presently relies  on
our  comprehension of the mechanisms with   which Supernovae Ia (SNIa)
emit  radiation (see  \citet{per99,  rie01})   and  of the    physical
processes that  produced   temperature fluctuations  in   the primeval
plasma (see \citet{lee01, deb02, hal02, spe06}.)

Even  if the ambitious task of   determining geometry and evolution of
the  universe as a whole, which  commenced in the 1930s, now-day shows
that the relativistic Friedman-Lema\^{\i}tre model passes impressively
demanding  checks, we are faced with  the  challenge of developing and
adding new  lines  of evidence  supporting   (or falsifying)  the {\it
concordance} model. Moreover, even if   we parameterize our   ignorance
about dark energy describing its nature only via  a simple equation of
state $w=p/\rho$, we only have  loose constraints on the precise value
of the w parameter or on its functional behavior.

In this  spirit we  focus  this  analysis on  possible   complementary
approaches    to determining   fundamental cosmological    parameters,
specifically on geometrical tests.

A whole arsenal of classical geometrical methods has been developed to
measure global properties of the universe. The  central feature of all
these  tests  is the attempt to  directly  probe the various operative
definitions    of  relativistic   distances   by  means   of   scaling
relationships in which an  observable  is expressed  as a function  of
redshift ($z$) and of the  fraction of critical density contributed by
all forms of matter and energy ($\Omega$).

The most remarkable among these classical  methods are the {\it Hubble
diagram} (or  magnitude-redshift relation $m=m(M,z,\Omega$)), the {\it
Angular          diameter     test}    (or   angle-redshift   relation
$\theta=\theta(L,z,\Omega$)), the {\it Hubble test} (or count-redshift
relation N=N(n,z,$\Omega$))   or the  {\it Alcock-Paczinsky  test} (or
deformation-redshift relation $\Delta z/ z \Delta \theta\equiv k= k(z,
\Omega$)). The common key idea is to constrain cosmological parameters
by measuring,  at various cosmic epochs, the   scaling of the apparent
values $m$, $\theta$, N, $k$ of some  reference standard in luminosity
(M),  size  (L),  density (n)   or   sphericity and  compare   them to
corresponding model predictions.

The  observational viability of these  theoretical strategies has been
remarkably proved by the Supernova Cosmology Project \citep{per99} and
the  High-z Supernova Team  \citep{rie01} in  the case  of  the Hubble
diagram. With a parallel strategy,  \citet{new02} recently showed that
a variant of the Hubble test ($N(z)$ test) can be in principle applied
to distant  optical clusters selected in  deep redshift survey such as
VVDS \citep{lef05} and DEEP2  \citep{dav00},  in order to measure  the
cosmic equation-of-state parameter $w$.

Unfortunately,  the   conceptually simple pure   geometrical tests of
world models,  devised     to anchor relativistic cosmology  to    an
observational basis,    have so  far   proved   to   be difficult   to
implement. This   is because the  most effective  way to constrain the
evolution of the cosmological metric  consists in probing deep regions
of   the universe      with a  primordial    class   of   cosmological
objects. Besides   the complex instrumental    technology this kind of
experiments  requires,   it becomes  difficult  at  high   redshift to
disentangle the effects  of   object evolution from the  signature  of
geometric evolution.

Since geometrical tests are by definition independent from predictions
of   theoretical models or  simulations,  as well  as from assumptions
about content, quality  and   distribution of matter in   the universe
(mass  fluctuations  statistics (e.g. Haiman,    Mohr \& Holder  2001,
Newman et al.  2001), galactic  biasing  (e.g. Marinoni et al.   1998,
Lahav et al.  2001,  Marinoni et  al.  2006), Halo   occupation models
(e.g. Berlind et al. 2001,  Marinoni \& Hudson 2002,  van den Bosch et
al.   2006)) it  is  of  paramount  importance  to  try  to  devise an
observational way to implement them. The technical maturity of the new
generation of   large  telescopes, multi  object  spectrographs, large
imaging   detectors and space   based  astronomical observatories will
allow these  tests to be  more effectively applied  in the near future
\citep{hut00}. In  this paper, we describe a  method to select a class
of  homologous  galaxies  that   are at  the  same  time  standard  in
luminosity and size,  that can be in  principle applied to data coming
from  the zCOSMOS   spectro-photometric   survey (Lilly et  al.  2006,
of the deep universe.

An observable relationship exists between the speed of rotation $V$ of
a spiral galaxy and  its metric radial  dimension $D$  as well as  its
total   luminosity    L   \citep{tul77,bot80}.    From   a theoretical
perspective, this set of scaling relations are expected and explicitly
predicted in the context of CDM models of galaxy formation \citep{MO}.
The  Tully-Fisher  relations for   diameter  and luminosity  have been
extensively  used in the local universe  to determine the distances to
galaxies and the value of  the Hubble constant.   We here suggest that
they may be used in  a cosmological context to  select in a physically
justified way, high redshift  standard rods since galaxies  having the
same    rotational  speed will  statistically     have the same narrow
distribution in physical sizes.

The picture gets complicated  by the fact  that the standard model  of
the universe implies some sort of evolution  in its constituents. In a
non  static, expanding universe, where  the scale  factor changes with
time,  we  expect various   galaxy properties, such  as  galaxy metric
dimensions, to be an explicit function of  redshift. In principle, one
may break this circular argument  between model and evolution with two
strategies: either by understanding the  effects of different standard
rod evolutionary patterns  on  cosmological parameters, or  by looking
for cosmological predictions  that  are independent from  the specific
form of the disc evolution function.

In this  paper, following the first approach,  we  study how different
disc  evolution functions  may   bias the  angular-diameter  test.  We
simulate the diameter-redshift experiment using the amount of data and
the realistic errors expected in   the context of the zCOSMOS  survey.
We  then   evaluate how  different disc   evolution functions   may be
unambiguously recognized from the data and  to what extent they affect
the estimated values of the various  cosmological parameters.  We also
explore the second approach and  show how cosmological information may
be  extracted, without any   knowledge about the particular functional
form  of the standard rod  evolution, only by requiring  as a prior an
estimate  of the upper limit value  for the relative disc evolution at
some reference redshift.

This paper is  set out as follow: in  \S  2 we review the  theoretical
basis of the angular diameter  test. In \S 3  we describe the proposed
strategy to select high redshift standard rods.  In \S 4 we digress on
how implement in practice the '$\theta-z$' test with zCOSMOS data, and
in  \S 5  we present  the zCOSMOS  expected statistical constraints on
cosmological  parameters. In   \S  6 we   discuss different   possible
approaches   with which   to  address the    problem of  standard  rod
evolution. Conclusions are drawn in \S 7.

\section{The Angular Diameter Test}

We investigate  the  possibility  of probing the  cosmological  metric
using the redshift  dependence of the  apparent angular  diameter of a
cosmic standard rod.  
What  gives this test special appeal   is the possibility of detecting
the ``{\it   cosmological lensing}"  effect, which  causes incremental
magnification
 of the apparent diameter of a fixed reference length.

Let's consider the transverse comoving distance \citep[see][]{hog99} 
 \begin{equation}
 r(z,\Omega_m,\Omega_Q,w)=\frac{c}{H_0\sqrt{\mid     \Omega_k   \mid}}
 S_{k}\big(\sqrt{\mid \Omega_k \mid}    \int_{0}^{z}  E(x)^{-1}dx\big)
 \label{one}
\end{equation} where \begin{equation} E(x)=[\Omega_{m}(1+x)^3+\Omega_Q
(1+x)^{3+3w}+\Omega_{k}(1+x)^2]^{1/2}  \label{due}  \end{equation} and
where $S_{-1}(y)=\sinh(y)$,  $S_{1}(y)=\sin(y)$,    $S_{0}(y)=y$ while
$\Omega_{k}=1-\Omega_m-\Omega_Q$.

An object with  linear  dimension $D$ at   a redshift $z$ has  thus an
observed angular diameter $\theta$

\begin{equation}    \theta(z,\vec{p})=\frac{D}{r(z,\vec{p})}     (1+z)
\label{angle} \end{equation}

which     depends  on  the general    set   of cosmological parameters
$\vec{p}=[\Omega_m,\Omega_Q,w]$ via   the   relativistic definition of
angular distance, $d_{A}=r(z,\vec{p})/(1+z)$.

This test may  be implemented without  requiring the knowledge  of the
present expansion rate of the universe (the dependence from the Hubble
constant  cancels out in    eq.  \ref{angle}). At variance,   although
characterized   by  a  smooth    and    diffuse nature, dark    energy
significantly  affect the dynamic of  the universe. From eq. \ref{one}
it is clear that the angular-diameter test depends  on the dark energy
component via the expansion rate evolution  $E(z)$.  The more negative
$w$, the  more accelerated  the expansion is  and  the smaller a fixed
standard rod will appear to an observer.

\begin{figure} \includegraphics[width=90mm,angle=0]{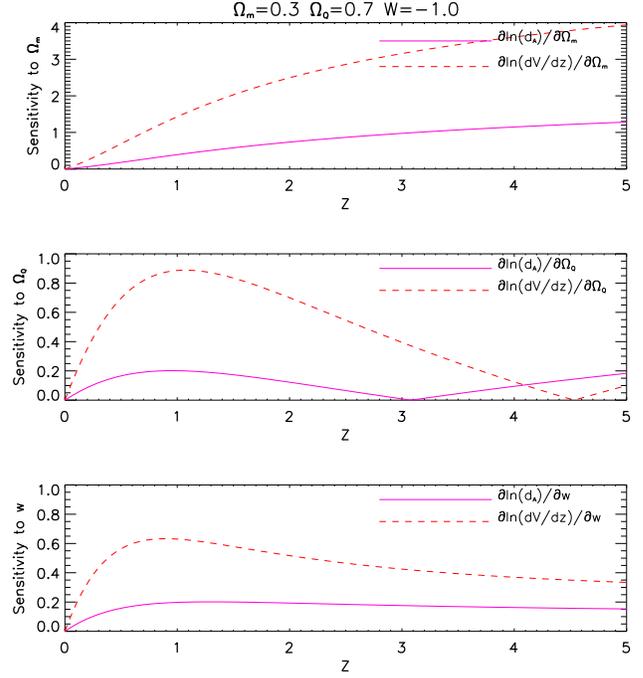} \caption{
The relative sensitivity of the  angular diameter distance ($d_A$) and
volume  element (dV/dz) to a    change in  the values of   $\Omega_m$,
$\Omega_Q$ and $w$. The partial derivatives  are computed with respect
to  the   position   ($\Omega_m=0.3, \Omega_{Q}=0.7,    w=-1)$ in  the
parameter space. } \label{fig1} \end{figure}

The efficiency of   different cosmological observables in probing  the
nature of space-time ultimately  depends upon their sensitivity to the
cosmological  parameters  $\Omega_m,   \Omega_Q,  w$.    The  relative
sensitivity of empirical cosmological tests  based  on the scaling  of
the angular  diameter distance ($d_{A}$)  and  of the  volume  element
($dV/dz=(c/H_0)(r^2/E(z)$)   is derived  in   Fig. \ref{fig1},  where we
assume    that   Poissonian errors   are    constant  in time  and  no
redshift-dependent      systematics     perturb      the  measurements
(e.g.  \citet{hut00}).     Since the  luminosity  distance   (i.e. the
distance inferred from  measurements of the  apparent  magnitude of an
object of  known absolute luminosity)  is defined as $d_L=(1+z)^2d_A$,
we  note that  the  angular diameter test   has the  same cosmological
discriminatory  power  as  the Hubble diagram.    The  upper panel  of
Fig. \ref{fig1} shows  that the sensitivity  of both $d_A$ and $dV/dz$
to  the     mean  mass    density parameter,   $\Omega_m$,   increases
monotonically as  a function of redshift.   This means that the deeper
the region of the universe surveyed, the more constrained the inferred
value of $\Omega_m$ is.

Conversely, the sensitivity of both empirical tests to a change in the
constant value of  $w$ peaks at redshift around  unity, and levels off
at redshifts  greater than $\sim 5$. The  reason for  this is that the
dark energy density  $\rho_Q$, which substantially contributes to  the
present-day value  of the expansion  rate was  negligible in the early
universe       ($\rho_Q/\rho_M\propto           (1+z)^{3w}$,       see
eq.  \ref{due}).

The fact that we are living in a special epoch, when two or more terms
in the  expansion rate equation make  comparable  contributions to the
present value of  $E(z)$, can be appreciated in  the central  panel of
Fig. \ref{fig1}. Because  each of the  terms in  eq.  \ref{due} varies
with cosmic time in a different way, there is  a redshift window where
the search for  $\Omega_{Q}$  is    less efficient  (i.e.    $2<z<4$).
Therefore one can maximize  the cosmological information which can  be
extracted from the classical tests of cosmology, and specifically from
the angular diameter  test, by devising observational programs probing
a large field of view in the redshift range $0 \leq z \leq 2$.

\section{The Standard Rod}

A  variety of standard  rod candidates have  been explored in previous
attempts  of implementing the angular diameter-redshift test:  
galaxies \citep{san72,  djo81},  clusters \citep{hic77,
bru78, pen97}, halo clustering \citep{coo01}.  Those methods failed to
yield conclusive evidence because the available redshifts were few and
local, and  the quality of the  imaging  data used in  the estimate of
sizes was poor.

Good quality size  measurements for high  redshift objects have become
available  for radio sources (e.g.  \citet{mil71, kap75}, and recently
several   authors \citep{kel93, wil98}     have reported   a  redshift
dependence  of radio source  angular sizes at  $0.5<z<3$, which is not
easily reconciled with  other recent measurements of  the cosmological
parameters  (but   see   Daly \&  Djorgovski   2004  for  results more
consistent with the concordance model.)

The radio source results may be affected by a variety of selection and
evolutionary effects, the lack of a  robust definition of size, and by
difficulties in assembling    a large,  homogeneous  sample of   radio
observations \citep{buc98, gur99}.

A common thread of weakness in all these  studies is that there are no
clear  criteria by which galaxies, clusters,  extended  radio lobes or
compact    radio  jets associated with  quasars     and AGNs should be
considered universal standard     rods. Moreover, lacking   any  local
calibration for the metric size of the standard  rod, the standard rod
dimension (parameter D  in eq. \ref{angle})  is often  considered as a
free fitting  parameter.  Since the  inferred cosmological  parameters
heavily depend on the assumed value for the object size \citep{lim02},
an  a   priori independent  statistical  study   of the  standard  rod
distribution properties is an imperative prerequisite.

\begin{figure} \includegraphics[width=90mm,angle=0]{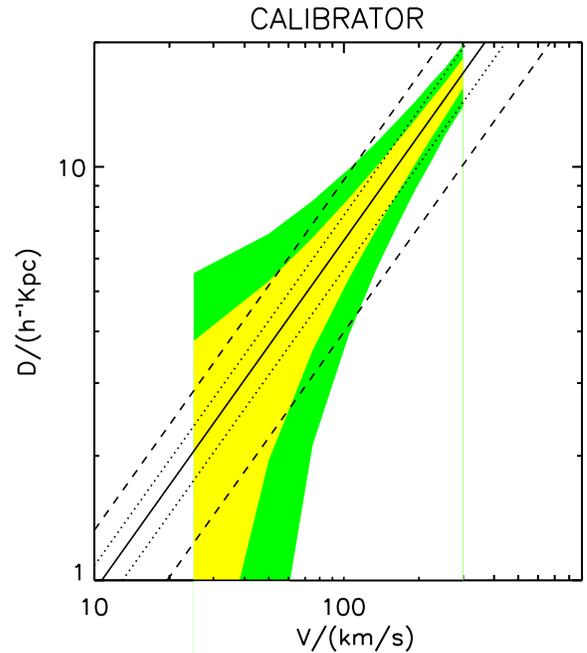} \caption{
The diameter  vs.  velocity  relationship  calibrated  in paper II  is
plotted  using a  black  line. Shadowed  regions  represent  the 1 and
2$\sigma$ uncertainties in  the calibrated relationship.  D represents
the corrected  (face on)  half  light diameter while the  velocity has
been measured using the $H_{\alpha}$ line.  Dotted lines represent the
upper  and lower relative  uncertainty  $[\sigma_{D}/D]_{int}=0.15$ in
diameters.  The    {\it conservative}    relative dispersion   in  the
relationship  assumed  in  this  study ($[\sigma_{D}/D]_{int}=0.4$) is
also presented  using dashed lines. With  this  conservative choice we
take into account  that sizes  and velocity  are measured  in the high
redshift  universe with  greater  uncertainties.  Within the  interval
centered at $V=200\pm 20$  km s$^{-1}$ (mean physical galaxy dimension
$\sim 10 \pm 1.5 h^{-1}$ kpc which roughly corresponds to an isophotal
diameter $D_{25} \sim 20  \pm 3 h^{-1}$  kpc), we will select galaxies
with    total absolute I  magnitude   $M_{I}-5\log h=-22.4$ (see paper
II). The velocity selected standard rods  at $z \sim  1$ are thus well
within the visibility window of  deep galaxy redshift surveys such  as
the VVDS \citep{lef05} or zCOSMOS \citep{lil06}.  For exemple the VVDS
is  flux limited at $I<24$   and selects objects  brighter then $M_{I}
\sim -20+5\log h$ at $z = 1$.  } \label{fig2} \end{figure}

We thus propose to use  information on the  kinematics of galaxies, as
encoded in their optical spectrum,
 a) to identify in an objective and  empirically justified way a class
    of objects behaving
as standard rods, and b) to measure the absolute value of the standard
rod length.  The basic  idea consists  in using  the velocity-diameter
relationship for disc galaxies \citep[e.g.][]{tul77, paperII} as
a cosmological   metric probe.  In Fig. \ref{fig2}   we plot the local
relationship we have derived  in  paper II of this series (\citep{paperII}) 
between half-light diameters
and    rotational velocities  inferred   using the  H$_{\alpha}\lambda
6563$\AA line. The sample used to calibrate the diameter-velocity 
relationship is the SFI++ sample described by Springob et al 2007.
Also shown  is the the amplitude  of the scatter in the
zero-point calibration of the standard rods.

In   the local universe, rotation    velocities can be estimated  from
either 21  cm  HI spectra  or from  the H$_{\alpha}\lambda  6563$\AA \
optical emission lines.  However,  the  H$_{\alpha}$ line  is  quickly
redshifted into the near-infrared  and cannot be used  in ground-based
optical galaxy redshift surveys at $z>0.4$, while HI is not detectable
much past $z=0.15$.  Only [OII]$\lambda  3727$\AA \ line widths can be
successfully used in optical surveys to infer the length of a standard
rod  dimension $D$ at $z\sim  1$.  Clearly one could obtain rotational
velocity information for   high   redshift objets by  observing    the
H$_{\alpha}$ line   with near-IR spectrographs.   However it  is  much
easier to  get large samples of kinematic  measurements using  OII and
multi-slit  devices, rather  than   get   sparser  samples  using    a
single-object, near-IR spectrograph.

A detailed study of the  kinematical information encoded in the  [OII]
line are presented in  Paper II.  In
that  paper we  have  explored the  degree of  correlation  of optical
$H_{\alpha}$ and [OII]  rotational velocity indicators,  i.e. how well
the  rotation velocities   extracted    from  these different    lines
compare.   Moreover,   we   have  derived  a   local diameter-velocity
relationship, and we have investigated the amplitude of the scatter in
the zero-point calibration of the standard rods.

We  finally note that the  present-day expansion rate sets the overall
size and time  scales for most other  observables in cosmology.  Thus,
if we hope to seriously constrain other  cosmological parameters it is
of  vital importance  either  to pin  down    its value or   to devise
$H_0$-independent cosmological tests. Note that, given the calibration
of the diameter--linewidth relation   in the form   $H_0 = f(V)$,  the
$\theta$-expression in equation \ref{angle} is effectively independent
of the value of the Hubble constant.

\section{Optimal Test Strategies}

In this section we outline the optimal observational strategy required
in order   to perform the proposed  test.  With the proposed selection
technique, the    photometric standard rod    $D$ is spectroscopically
selected and the    sample  is therefore free from     luminosity-size
selection   effects, that is from the   well known  tendency to select
brighter    and   bigger  objects    at   higher  redshifts (Malmquist
bias) in flux-limited samples.  However it  is  crucial  that a  large sample  
of   spectra be
collected,  in   order to obtain $\sqrt{N}$   gain  over the intrinsic
scatter in the calibrated $V$(OII)-diameter relationship.

\subsection{Galaxy sizes measurement}

Since galaxies   do not have  sharp edges,  their angular  diameter is
usually  defined  in   terms of  isophotal   magnitudes. However since
surface brightness is  not constant  with   distance, the  success  in
performing  the experiment revolves   around the use of metric  rather
than isophotal galaxy diameters \citep{san95}.

A suitable way to measure  the photometric parameter $\theta$, without
making any  {\it   a-priori}  assumption  about  cosmological  models,
consists in adopting as the standard scale length estimator either the
half-light radius of the galaxy  or  the $\eta$-function of  Petrosian
(1976).  The Petrosian radius is implicitly defined as

\begin{equation}         \eta   (\theta)=\frac{\langle \mu    (\theta)
\rangle}{\mu(\theta)},
 \end{equation}

i.e. as the radius $\theta$ at which
the surface brightness averaged inside $\theta$ is a predefined factor
$\eta$ larger than the local surface brightness at $\theta$ itself.

Both   these   size    indicators   are independent    of   K-correction,  
dust  absorption, luminosity evolution (provided 
the evolutionary change of surface brightness is independent of radius), 
wave-band  used (if there is no color gradient)  and source  light
profile \citep{djo81}.

\subsection{Standard rods optimal selection}

The choice of the objects for which the velocity parameter $V$ and the
metric size   is  to   be measured   is  a    compromise  between  the
observational need  of detecting  high  signal-to-noise  spectral  and
photometric features ({\it  i.e.} selecting high luminosity  and large
objects)  and the  requirement of  sampling the velocity  distribution
function ($   n(V)dV \sim V^{-4}$   for galaxy-scale halos)  within an
interval where the rotator density is substantial.

Given  the estimated  source  of errors  (see next  section),  and the
requirement of determining both $\Omega_{Q}$ and  $w$ with a precision
of  $20\%$, we find, guided  by semi  analytical models predicting the
redshift distribution of  rotators \citep[i.e.][]{nar88, new00},  that
an optimal choice are $V=200$ km s$^{-1}$ rotators.

In  particular, as  shown  in Paper   II,  the I  band  characteristic
absolute  magnitude of the $V=200$ km  s$^{-1}$ objects is $M_{I} \sim
-22.4+5\log h_{70}$
i.e. well above the visibility threshold of flux-limited surveys
such as  zCOSMOS or VVDS (as  an example, for $I_{ab}=24$ the limiting
magnitude  of the  VVDS, one  obtains  that $M_{I} \lesssim  -21+5\log
h_{70}$  at $z=1.5$).   Therefore,  with  this  velocity  choice,  the
selected  standards do not  suffer  from any  Malmquist bias (i.e. any
effect which favors the systematic selection  of the brighter tails of
a luminosity distribution at progressively higher redshifts).

\subsection{The zCOSMOS Potential}

In \S  2  we have emphasized  that  an optimal strategy   to study the
expansion history  of   the universe consists  in probing,    with the
angular diameter test,  the $0 \leq  z \leq 2$  interval.  However, in
order to collect a large and minimally biased  sample of standard rods
over such a wide redshift baseline, we  need the joint availability of
high quality images and high resolution multi-object spectra.

Space images allow  a  better   determination of galaxy     structural
parameters (sizes, luminosity, surface brightness, inclination etc) at
high  redshift.   In particular, the   ACS camera of  the Hubble Space
Telescope can survey large sky  regions with the  key advantages of  a
space-based experiment: diffraction limited images no seeing blurring,
and very deep  photometry. Ground multi-object spectrographs operating
in   high resolution mode  allow   a  better characterisation  of  the
gravitational   potential-well   of  galaxies, facilitating  the  fast
acquisition of  large  samples of standard rods.  For  example, in the
spectroscopic resolution mode R=2500 (1"slits), the VIMOS spectrograph
(Le F\`evre et al. 2003) allows  to resolve the internal kinematics of
galaxies via their rotation curve or  line-width.  Note that the VIMOS
slitlets  can be  tilted and  aligned   along the  major  axis of  the
galaxies in order to remove a source  of potentially significant error
in the estimate of the rotation velocities.

\begin{figure*}       \includegraphics[width=134mm,angle=-90]{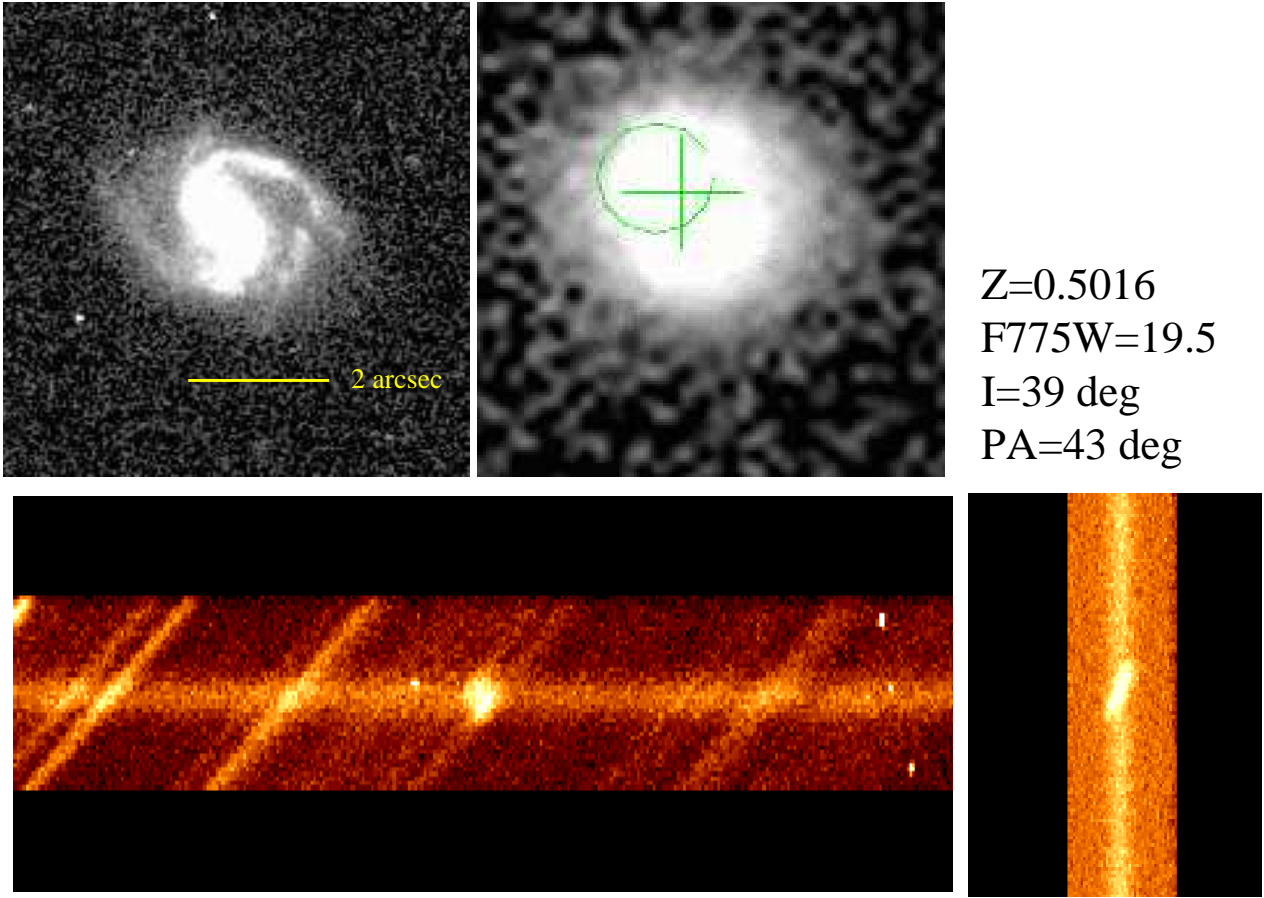}
\caption{  {    \em Upper:}  Public  release  image     of the  galaxy
$\alpha=53.1874858$,   $\delta=-27.910975$  at redshift  0.5016, taken
with the filter F775W  (nearly I band) of  the ACS camera by the GOODs
program.  For comparison the same galaxy  as imaged  in the EIS survey
with the WFI camera   at the ESO 2.2mt telescope   at La Silla. {  \em
Botton left:} raw spectrum of the galaxy taken by VIMOS at the VLT-UT3
telescope  with  a  total exposure time   of  90  min  and a  spectral
resolution R=2500.  Slitlets have been  tilted according to  the major
axis orientation (Position Angle = 43$^{\circ}$). { \em Botton right:}
final processed spectrum showing the  rotation curve as traced by  the
Hb ($\lambda 4859$\AA) line.  } \label{fig4} \end{figure*}

These    observational requirements  are   mandatory for  a succesfull
implementation  of the   proposed cosmological  probe.  The  practical
feasibility of   the   strategy is   graphically illustrated  in   Fig
\ref{fig4}.   In this figure  we show  a high resolution  spectra of a
galaxy  at  z=0.5016 obtained with  a total   exposure of  90 min with
VIMOS. The ground and space (ACS) images of the  galaxy are also shown
for comparison.

Interestingly, a large sample of rotators can  be quickly assembled by
the  currently underway zCOSMOS deep redshift   survey, which uses the
VIMOS multi-object spectrograph at the VLT to target galaxies with ACS
photometry  in the 2 sq. degree  COSMOS field (Scoville et al. 2003)).
In  principle, one can measure  in high resolution modality R=2500 the
line-widths   of  [OII]$\lambda  3727$\AA~  up  to   $z \sim   1.4$ by
re-targeting objects for which the redshift  has already been measured
in  the low-resolution   (R=600)  zCOSMOS survey.  Since the  spectral
interval covered in high  resolution mode is limited to $\sim$2000\AA,
this    allows us to  determine    the optimal telescope and  slitlets
position angles  in order to maximize  the number of spectra whose OII
emission lines  fall onto the CCDs.  With this fast follow-up strategy
one could be able to target about  $200$ rotators per pointing with an
exposure time of 4 hours down to $I_{ab}=24$.

In the  following we will  use the zCOSMOS sample  as  a test  case to
assess the performances of the proposed cosmological test.

\begin{figure*}       \includegraphics[width=114mm,angle=-90]{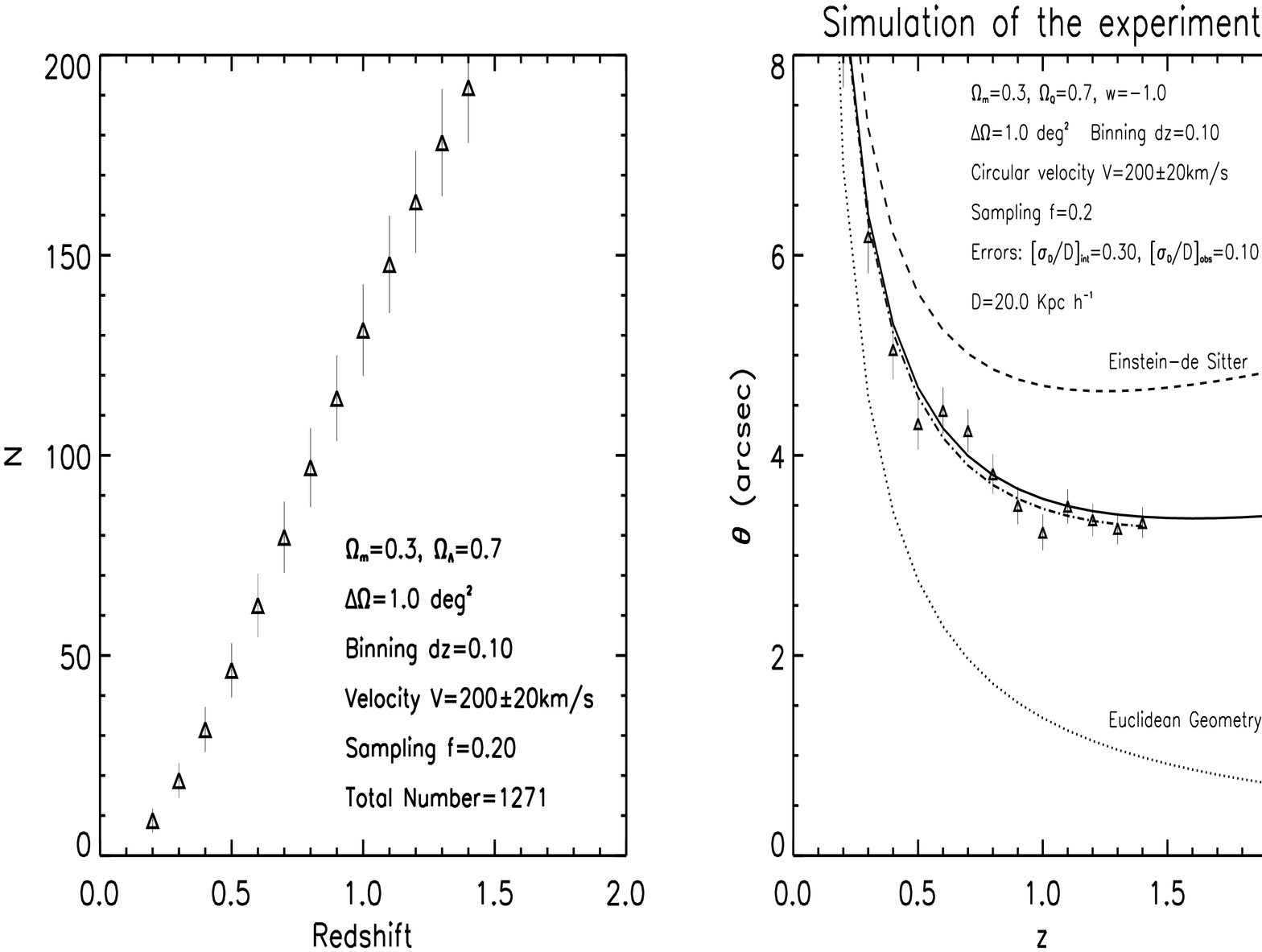}
\caption{ { \em Left:} redshift evolution of the differential comoving
number density of halos with a circular  velocity of $200$ km s$^{-1}$
computed according to the  prescriptions of \citet{new00} in  the case
of  a flat  cosmological model having  $\Omega_m  =  0.3$ today  and a
cosmological constant.   Only a fraction f=0.2  of the total predicted
abundance of halos ({\em i.e.} $\sim  1300$ objects per square degree)
is conservatively supposed to  give line-width information useful  for
the angular-diameter test.  {\em Right:}  simulation of the  predicted
scatter   expected to   affect  the angular  diameter-redshift diagram
should  in  principle  achieve with  the  angular  diameter  test. The
simulation is performed assuming the sample is composed by $\sim 1300$
rotators  with $V=200$    km s$^{-1}$  and  that  a  flat   model with
parameters $[\Omega_m=0.3, \Omega_Q=0.7, w=-1]$ is the true underlying
cosmological framework.  The circular velocity has been converted into
an estimate of   the galaxy diameter  ($D_v=20  \; kpc$)  by using the
velocity-diameter template   calibrated by  \citet{bot80}.   The  {\it
worst-case} scenario  ($[\sigma_D/D]_{int}=40\% $)  is presented.  The
solid  line  visualizes   the  underlying   input  cosmological  model
$\theta^{\Lambda CDM}(z)$, while triangles are drawn from the expected
Poissonian fluctuations.  The  dot-dashed line represents the expected
scaling  of  the angular diameter  in  our best recovered cosmological
solution.  The  dashed  and  the  dotted  lines  represent the angular
scaling in a  Einstein-de Sitter and Euclidean (non-expanding type
cosmology with zero curvature)  geometry respectively.
} \label{fig5} \end{figure*}

\section{Constraints on Cosmological Parameters}

In this section we present in detail some merits and advantages of the
proposed approach for constraining the value of the fundamental set of
cosmological parameters. We evaluate the potential of the test, within
the zCOSMOS operational  specifications,  in placing  constraints  not
only on the simplest models of the universe, which include only matter
and a cosmological   constant, but also on so-called   ``quintessence"
models \citep{tur97, new02}, For  the purposes of this study, we assume
in this  section  $w$  to be  constant  in  time up to   the  redshift
investigated $z \sim 1.4$.

We assume that a $V$(OII)-diameter relation can be locally calibrated,
and that the diameter of $V=200$ km s$^{-1}$  rotators may be inferred
with  a       worst(/optimal)            case    relative        error
$[\sigma_D/D]_{int}=40\%(/15\%)$ (see \S 3  and Fig  \ref{fig2}.)  The
main contributions to this error figure are uncertainties in measuring
linewidths,  galaxy inclinations and    the intrinsic scatter  of  the
empirical relation  itself. We then   linearly combine this  intrinsic
scatter with the uncertainties  with which the Petrosian or half-light
radius may be determined  from the ACS  photometry.  The error  on the
half   light   radii  is  about   $\sigma_{\theta}=\pm   0.04"$ almost
independent of  galaxy sizes in   the range 0.1-1" (S.  Gwyn,  private
communication).

We  consider as the observable of  the experiment the logarithm of the
angle subtended by a standard rod. We use the logarithms of the angles
rather than  angles themselves  because   we assume that   the  object
magnitudes, rather   than diameters, are normally   distributed around
some mean  value. Moreover, in this  way the galactic diameter becomes
an additive  parameter, whose fitted value (when  a z=0 calibration is
not available)  does not distort    the  cosmological shape  of    the
$\theta(z)$ function.

The observed values  $\log \theta$ are  randomly simulated around  the
theoretical value  $\log \theta^{r}$ (cfr.  eq. \ref{angle}) using the
standard deviation given by

\begin{equation}                                                \sigma
=\Big[\Big(\frac{\sigma_D}{D}\Big)^{2}_{int}+\Big(\frac{\sigma_{\theta}}{\theta}\Big)^{2}_{obs}\Big]^{1/2}
\label{err} \end{equation}

For the purposes of this  section, $\theta^{r}$ is computed assuming a
flat, vacumm dominated
 cosmology with parameters  $\Omega_m=0.3, \Omega_{Q}=0.7$, and $w=-1$
 as reference model.
The confidence with  which these parameters  are  constrained by noisy
data is evaluated using the $\chi^2$ statistic

\begin{equation}     \chi^2=\sum_{i}  \frac{[\log    \theta_{i}-  \log
\theta^{th}_{i}(z, \vec{p})]^2}{\sigma^2}, \label{chi} \end{equation}

where $\theta^{th}$ is given by eq. \ref{angle}.

We  also derive the expected redshift  distribution of galaxies having
circular velocity in the $180 \leq V \leq  220 $ km s$^{-1}$ range, in
various  cosmological scenarios within the  framework  of the Press \&
Schechter  formalism \citep{nar88, new00}.   We note  that due to  the
correlation between circular  velocity and luminosity, these  galaxies
could  be observed to  the maximum depth ($  z \sim 1.4$) out to which
the  OII is  within the visibility   window of  VIMOS.    We take into
account the uncertainties  in the  semi-analytic predictions and   our
ignorance about the fraction of discs to be observed  that will have a
spectroscopically  resolved     [OII]$\lambda 3727$\AA~      line,  by
multiplying the  calculated halo  density by a  ``conservative" factor
f=0.2. Using the VIMOS R=2500 resolution mode data,  we thus expect to
be  able to   implement our  angular-diameter  research  program using
nearly 1300 standard  rods per square  degree, which is what has  been
simulated (see Fig. \ref{fig5}.)

\begin{figure*}         \includegraphics[width=150mm,angle=0]{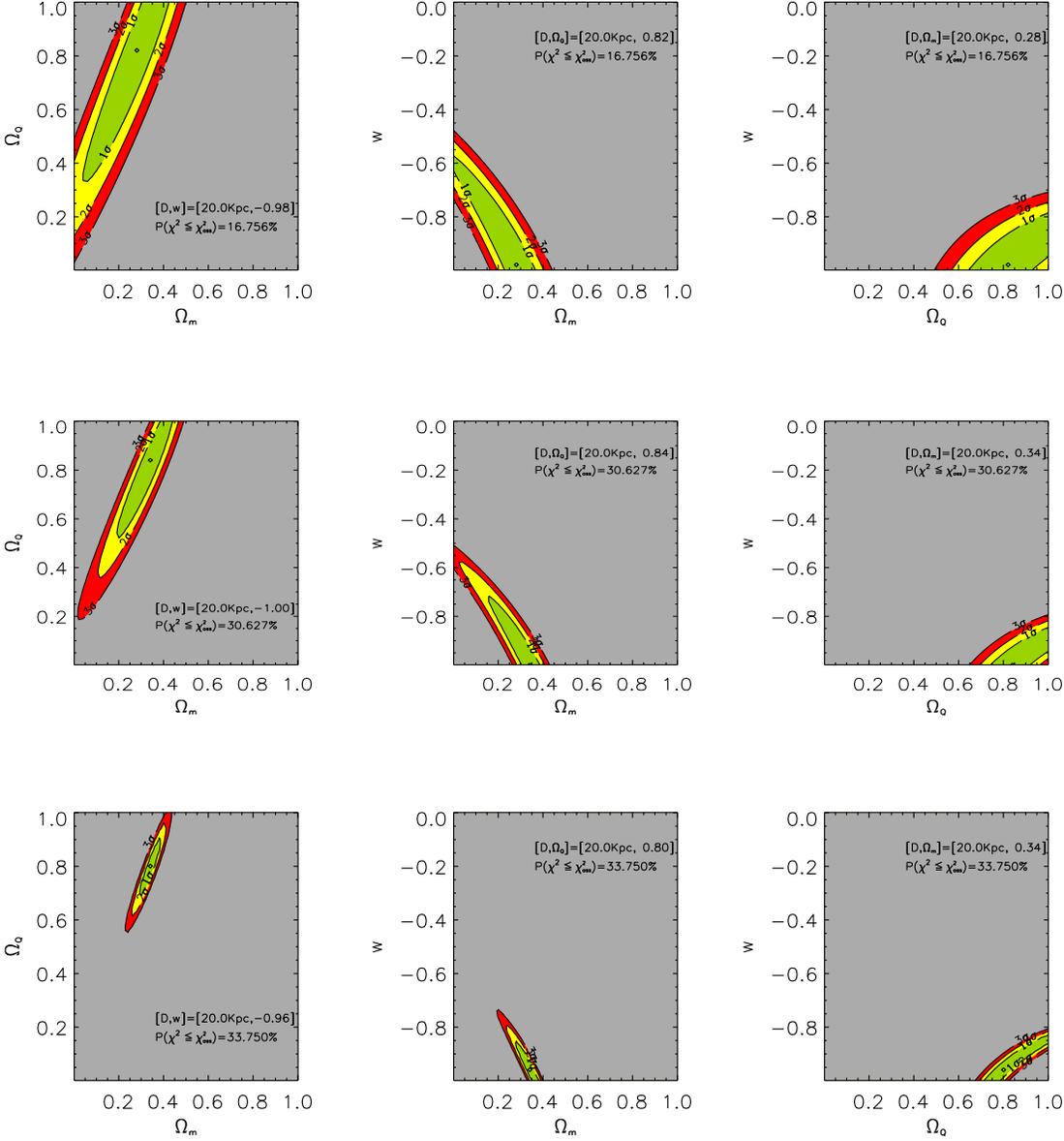}
\caption{Predicted  1,2,and  3$\sigma$ confidence level  contours  for
application of the angular-diameter test. The likelihood contours have
been derived  by adopting  a $\Lambda$-cosmology $[\Omega_m, \Omega_Q,
w]=[0.3,0.7,-1]$  as the fiducial model   we want  to recover, and  by
conservatively assuming  one can obtain useful line-widths information
for a sample of galaxies    having the redshift distribution and   the
Poissonian    diameter  fluctuations  simulated   in  Fig. \ref{fig5}.
Confidence   contours are projected   onto  various  2D-planes of  the
$[\Omega_m, \Omega_Q, w]$ parameter space, and the jointly best fitted
value   along  the projection  axis,   together  with the  statistical
significance of  the fit, are reported in  the insets. Note the strong
complementarity  of  the   confidence  region   orientation which   is
orthogonal to  the  degeneration axis  of the  CMB measurements.  {\it
Top:)} constraints derived   assuming  that one  might  survey  only 1
square degree   of sky and  that   the  $V$(OII)-diameter relation  is
locally calibrated with a 40\%  of relative scatter in diameter.  {\it
Center:)}  as  before     but  assuming  a   scatter  of   15$\%$   in
diameters. {\it  Bottom:)}  confidence  contours  for a   survey of 16
deg$^2$ (which corresponds to the full area surveyed by VIMOS-VLT Deep
Survey)  assuming  a  template   V(OII)-diameter relationship  with  a
scatter of 30$\%$ in diameters.  } \label{fig6} \end{figure*}

Since we do not  make {\it a-priori}  assumptions about any parameter,
and in particular we do not assume a flat cosmology, results should be
distributed as a $\chi_{\nu}^2$ with $\nu=3$ degrees of freedom, which
can be directly translated   into statistical confidence  contours, as
presented in  Fig.  \ref{fig6}.  This  figure shows  that even without
assuming   a  flat    cosmology  as   a  prior     and  considering  a
diameter-linewidth  relationship  with a  40\% scatter,  by targetting
$\sim 1300$ rotators   we can directly  infer the  presence of a  dark
energy component with a confidence level better than $3 \sigma$.
 At the same time, its equation of state can  be constrained to better
 than
20\%  ($\sim  10\%$  if  the  the  diameter-linewidth  relationship is
calibrated with a $\sim 15\%$ relative precision).

\section{Standard Rod Evolution}

The   previous  analysis shows that  the   angular diameter test, when
performed  using     fast   high  resolution   follow-up   of  zCOSMOS
spectroscopic targets  may be used as  a  promising additional tool to
explore the  cosmological parameter space and  directly measure a dark
energy  component.  However, the  impact of any standard rod evolution
on these results needs to be carefully examined.

\begin{figure*}         \includegraphics[width=164mm,angle=0]{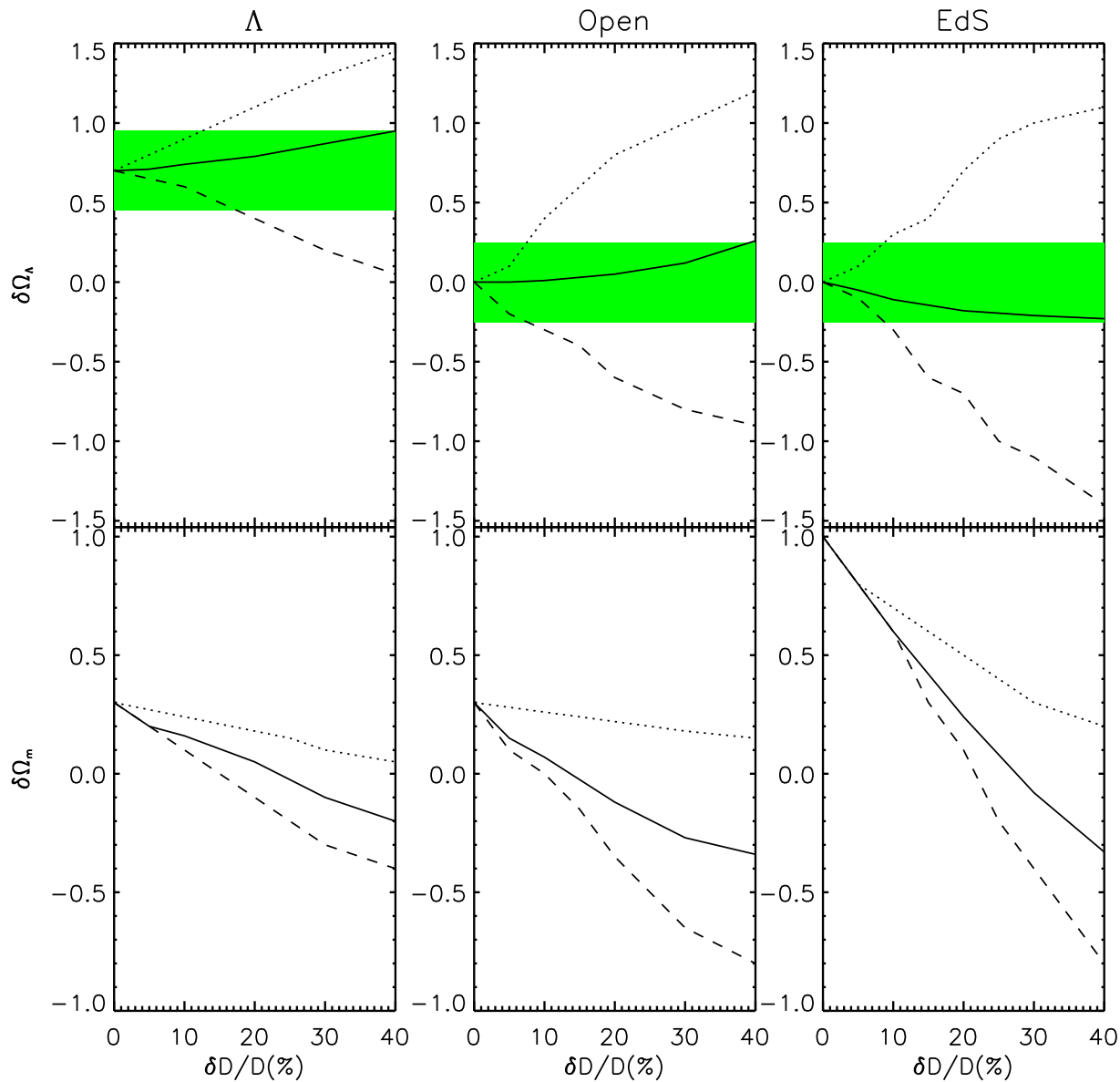}
\caption{Best fitting cosmological parameters inferred by applying the
angular  diameter   test to data affected    by  evolution. The output
(biased) estimates of $\Omega_m$ and $\Omega_{\Lambda}$ are plotted as
a function  of   the  relative diameter evolution   for  the following
evolutionary models: linear  (solid  line), square-root (dotted  line)
and  quadratic   evolution  (dashed line).   The   biasing pattern  is
evaluated    for  three  different    fiducial   cosmologies: a  flat,
$\Lambda$-dominated cosmology ($\Omega_{\Lambda}=0.7$, {\it left}),  a
low-density open cosmology  ($\Omega_m=0.3$, {\it center}) and a flat,
matter-dominated model  ($\Omega_{m}=1$, {\it  right}). } \label{fig7}
\end{figure*}

\begin{figure*}         \includegraphics[width=164mm,angle=0]{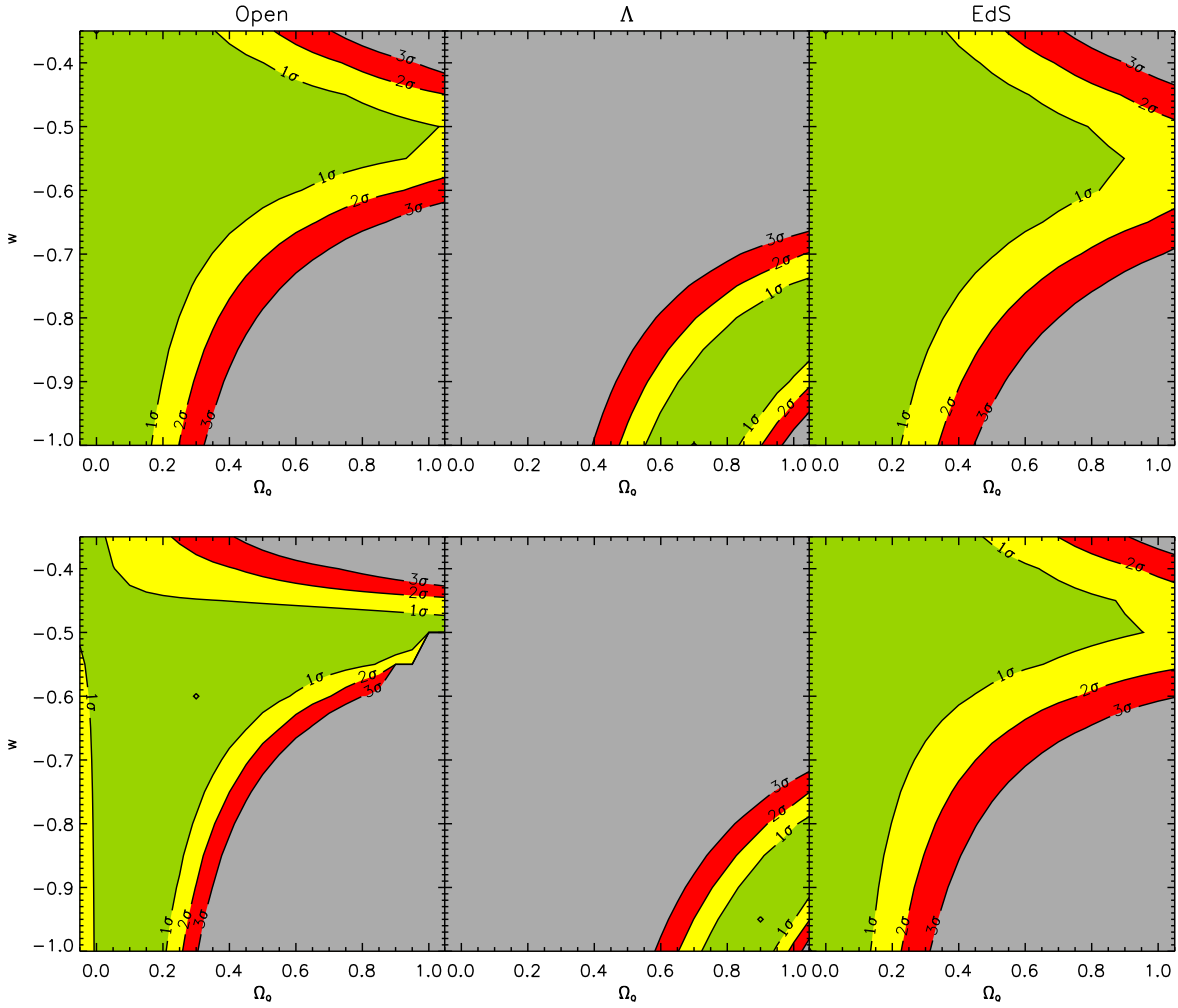}
\caption{   1,2,and  3$\sigma$  confidence    level  contours  in  the
[$\Omega_Q,w$] plane computed by applying the angular-diameter test to
data unaffected ({\it upper panel}) and affected by diameter evolution
({\it  lower  panel}). We consider a  linear  model for disc evolution
normalized by  assuming that discs were smaller  by $30\%$ at $z=1.5$,
and a nominal relative error in the standard rod measures of $5\%$ per
redshift bin  (see  discussion  in  \S  6.1).   The  effects  of  disc
evolution onto  cosmological parameter estimation  are compared to the
evolution-free  case for three  different  fiducial cosmologies: a low
matter density open  cosmology ($\Omega_m=0.3$,  {\it left}) a   flat,
$\Lambda$-dominated  cosmology ($\Omega_{\Lambda}=0.7$, {\it center}),
and  an Einstein-de  Sitter model  ($\Omega_{m}=1$,   {\it right}).  } 
\label{fig8} \end{figure*}

First of all, we may note that the expected variation with cosmic time
of  the total  galaxy  luminosity due    to evolution  in  its stellar
component  does not affect the metric  definition of angular diameters
unless  this  luminosity change depends on   radius (see paper III
of this series (marinoni et al. 2007) for a detailed analysis of this issue).
Moreover, we can
check   each  galaxy spectrum   or image  for peculiarities indicating
possible evolution or instability  of the standard   rod which may  be
induced  by  environmental effects,   interactions  or excess  of star
formation.

Any possible size evolution of the standard rod needs to be taken into
account  next.  Interestingly, it has been  shown by different authors
that large discs  in high redshift   samples evolve much less in  size
than in luminosity in the redshift range $0<z<1$.  Recent studies show
that  the  amount  of evolution  to  z$\sim$1 appears   to be somewhat
smaller than expected: disc  sizes  at $z  \sim$ 1 are  typically only
slightly  smaller  than  sizes  measured locally  \citep{tak99, fab01,
nel02, tot02}.  This  is also theoretically predicted by
simulations; \citet{boi01}, for example,  show that large  discs ({\it
i.e.} fast rotators)  should have basically completed their  evolution
already  by  $z \sim   1$ and undergo  very  little  increase  in size
afterwards.  Infall  models  (e.g.,  Chiappini, Matteucci, \&  Gratton
1997, Ferguson \&  Clarke 2001, Bouwens   \& Silk2002) also predict  a
mild disk size evolution. Disk sizes at $z \sim$ 1 in these models are
typically only 20\% smaller than at $z = 0$.

\subsection{Analysis of the biases introduced by evolution}

Even if literature evidences are encouraging, we have to be aware that
even a small amount of evolution may introduce artificial features and
bias the reliability of the cosmological inferences.

In   this  section  we   directly address  this  issue  by considering
different  evolutionary patterns   for   the  standard rods,  and   by
analyzing to  what level the   simulated  true cosmological model  may
still be  correctly inferred using evolved data.   In  other words, we
investigate  how  different  disc  evolutionary histories  affect  the
determination of cosmological parameters by answering to the following
three questions:

{\it a)} is   there a feature  that  may be  used to discriminate  the
presence of evolution in the data?

{\it  b)}  which cosmological  parameter    is more sensitive  to  the
eventual presence of  disc evolution?, and  in particular what are the
effects of evolution on the value of  the inferred dark energy density
parameter $\Omega_{Q}$?

{\it c)}  is there a particular  evolutionary  scenario for  which the
inferred values of $\Omega_{Q}$ and w are minimally biased?

For the  purposes  of  this analysis,   we consider for    the angular
  diameter-redshift test the baseline ($0.1 \leq z \leq 1.4$) divided in
  bins of width $dz=0.1$ and assume a relative scatter in the mean size per bin 
  of $5\%$.  This scatter nearly corresponds to that expected for  a sample of 1300
  rotators (with  $0<z<1.4$ and  $dz=0.1$)  whose diameters  are individually (and locally)
  calibrated with  a $40\%$ precision. We then  select a given fiducial
cosmology  (input cosmology), apply  an   arbitrary evolution  to  the
standard rods, and   then fit  the evolved   data with  the  unevolved
theoretical prediction given in eq. \ref{angle} in order to obtain the
best  fitting (biased) output  cosmology and the associated confidence
levels  contours.  We decide  that the best fitting cosmological model
offers a good  fit to the evolved data  if the probability of a  worse
$\chi^2$ is smaller that $5\%$ (i.e. $P(\chi^2>\chi^{2}_{obs})<0.05$).

We adopt three   different parameterizations to  describe an  eventual
redshift  evolution of  the velocity  selected  sample of galaxy discs
$D_v$:     a   {\it    late-epoch}     evolutionary scenario  $(\Delta
D/D \equiv (D_v(z)-D_v(0))/D_v(0)=-|\delta_1| \sqrt{z})$ where  most of the  evolution
is expected to  happen  at low redshifts   and levels off  at  greater
distances ($\delta_1$ is the relative disc evolution at z=1), 
a {\it linear} evolutionary scenario $(\Delta D/D=-|\delta_1| z)$
without  any preferred scale  where major  evolutionary phenomena take
place (i.e. the gradient of the evolution  is nearly constant), and an
{\it early-epoch}  evolution scenario $(\Delta  D/D=-|\delta_1| z^2)$  where
most of the evolution is expected to happen at high redshift.

We note  that for  modest  disc evolution, the
linear parameterization satisfactorily describes the  whole class of evolutionary
models whose  series   expansion  may be   linearly represented   (for
example, the  hyperbolic model $(D_v(z)=D_v(0)/(1+|\delta_1| z))$).  For $z<<1$
it     also     represents  fairly   well     the   exponential  model
$(D_v(z)=D_v(0)(1-z)^{\delta})$.  Moreover, the linear model is the favored
scenario for disc size evolution  at least at low redshift  ($\lesssim
1.5$) as   predicted  by  simulations   \citep[e.g.][]{MO,bou00}.

First, let's assume that $w=-1$ and that  the dark energy behaves like
Einstein's  cosmological  constant.  In  Fig.  \ref{fig7} we  consider
three   different input    fiducial   cosmological models     (a  flat
$\Lambda$-dominated universe  ($\Omega_{\Lambda}=0.7$), an  open model
($\Omega_m=0.3$)  and an Einstein-de  Sitter  universe ($\Omega_m=1$))
and show the characteristic pattern traced  by the best fitting output
values ($\Omega_m$,   $\Omega_{\Lambda}$)  inferred  by  applying  the
angular diameter test to data affected by evolution.  A common feature
of  all the various evolutionary schemes  considered is that the value
of $\Omega_m$  is  systematically underestimated with  respect  to its
true input value:   the stronger the   evolution in diameter  and  the
smaller $\Omega_m$  will be,  irrespectively   of the particular  disc
evolutionary model  considered.  Since many  independent  observations
consistently indicate the existence of a  lower bound for the value of
the normalized matter  density  ($\Omega_m \gtrsim 0.2$), we  can thus
use this parameter as a sensitive indicator of evolution.

Once the presence of evolution is recognized, the remaining problem is
to  determine   the  level  of bias   introduced in   the dark  energy
determination.   If  the  gradient  of  the disk   evolution  function
increases   with  redshift   ({\it  quadratic}  evolution),  then  the
estimates of $\Omega_{\Lambda}$  are  systematically  biased  low. The
contrary happens if the evolutionary gradient  decreases as a function
of  look-back time ({\it square  root}  model).  If the disc evolution
rate is constant ({\it linear} model),  then even if discs are smaller
by a  factor as large as  $40 \%$ at $z=1.5$  the estimate of the dark
energy parameter is only minimally biased.  The net effect of a linear
evolution  is     to   approximately   shift     the    best   fitting
$\Omega_{\Lambda}$ value  in  a direction  parallel to  the $\Omega_m$
axis in the [$\Omega_{m},\Omega_{\Lambda}$] plane.

More generally, by linearly evolving disc sizes so that they are up to
$40\%$ smaller at z=1.5, and by simulating the apparent angle observed
in any  arbitrary  cosmological model with  matter  and energy density
parameters in the  range  $0<\Omega_m<1$ and $0 \leq  \Omega_{\Lambda}
\leq  1$ ($w=-1$)  we  conclude   that the  maximum  deviation of  the
inferred biased value of $\Omega_{\Lambda}$ from its true input value,
is  limited   to  be  $|max(\delta   \Omega_{\Lambda})|\lesssim  0.2$,
whatever the true  input value of the energy  density parameter is. In
other  terms, in the  particular case of a  linear and substantial ($<
40\%$ at z=1.5)  evolution of galaxy discs, the  central  value of the
dark energy parameter is minimally biased for any fiducial input model
with $0\leq \Omega_{m}<1$ and $0\leq \Omega_{\Lambda}\leq 1$).

We  have shown   that  the  presence  of  evolution is   unambiguously
indicated  by  the "unphysical" best fitting    value of the parameter
$\Omega_m$. We now investigate the amplitude  of the biases induced by
disc evolution  in the  [$\Omega_{Q}, w$]  plane.   We assume for this
purpose that  $w$ is free to  vary in the  range $-1\leq w \leq -1/3$,
which means assuming that the late  epoch acceleration of the universe
might be explained in terms of a slow rolling scalar field.

We first consider a situation where the disc size evolution is modest,
and  could be represented   by  any of  the three  models  considered.
Whatever  the    mild evolution model    considered  (less then $15\%$
evolution   from  z=1.5)   and assuming  a    scatter  in the  angular
diameter-redshift diagram of  5\% in each redshift  bin, we  find that
the input values   of $\Omega_{Q}$ and  $w$ are   contained within the
1$\sigma$ biased confidence contour derived from the evolved data.

Fig.  \ref{fig8}  shows the 1,  2  and $3  \sigma$ "biased" confidence
contours obtained by     fitting  with eq.   \ref{angle}  a  simulated
angular-diameter redshift  diagram  in which discs  have been linearly
evolved. We  show  that, {\em if  disc  evolution depends  linearly on
redshift and  causes galaxy dimensions to  be up  to $30\%$ smaller at
$z=1.5$}, the true  fiducial input values  of $\Omega_{Q}$ and $w$ are
still within  1$\sigma$ of the biased  confidence contours inferred in
presence  of a standard rod evolution   (and a scatter  in the angular
diameter-redshift relation as  low as 5\% in  each  redshift bin).  We
have tested that these conclusions hold  true for every fiducial input
cosmology with parameters in the range $0\leq \Omega_m \leq 1$, $0\leq
\Omega_{Q} \leq 1$ and $-1 \leq w \leq -1/3$.

Thus, if, disc   evolution   is linear (as  predicted   by theoretical
models) and  substantial (up to $\sim 30\%$  at $z=1.5$), or arbitrary
and  mild (up to   $\sim 15\%$ at  $z=1.5$), then  in  both  cases the
angular  diameter   test reduces from  a   test  of the whole   set of
cosmological parameters, to a direct and fully geometrical test of the
parameters subset  ($\Omega_Q$,$w$).   For    example, in  a   minimal
approach, the angular diameter test could be used  to test in a purely
geometrical   way  the  null  hypothesis  that   ``{\em a dark  energy
component   with a constant    equation   of state parameter $w$    is
dominating the  present day dynamics of  the universe}".  Moreover, as
Fig. \ref{fig8}  shows, an universe   dominated by dark energy  may be
satisfactorily   discriminated  from   a  matter  dominated   universe
($\Omega_Q=0$).

In the evolutionary pictures considered,  galaxy discs are supposed to
decrease  monotonically in size in  the past. Since the sensitivity of
the test to changes in the linear diameter D is described by a growing
monotonic  function in the redshift interval  [0,1.4], one may hope to
test cosmology in a  way which is less  dependent on systematic biases
by limiting the sample at $z \leq 1$.  However, by doing this we would
halve the  number of standard  rods available for the  analysis ($\sim
600$ with respect to the original $\sim 1300$). The test efficiency in
constraining  cosmological   parameters   would     consequently    be
significantly degraded.

\subsection{The Hubble-Diagram using galaxies}

The same velocity criterion that allows the selection of standard rods
also allows the selection of  a sample of  standard candles. Using the
same set of tracers,  the  Tully  \& Fisher relationships   connecting
galactic rotation  velocities   to luminosities and   sizes offer  the
interesting  possibility  of  implementing  two different cosmological
tests, the angular-diameter test and the Hubble diagram.

Thus, we  have analyzed  how the effects  of luminosity  evolution may
bias the estimation of   cosmological parameters in   the case of  the
Hubble  diagram.   Assuming that   the absolute luminosity  L   of the
standard candle  increases as a  function of redshift according to the
square   root, linear and  quadratic    scenarios, then  the value  of
$\Omega_m$ is systematically overestimated. The value of $\Omega_m$ is
biased in an opposite sense with  respect to the angular diameter test
(see  Fig. \ref{fig7}).   Therefore,  it  is  less  straightforward to
discriminate the eventual presence of evolution in the standard candle
on  the  basis of the  simple requirement  that  any "physical" matter
density  parameter  is characterized by  a  positive lower  bound. The
different $\Omega_m$ shifts   (with  respect to the  fiducial   value)
observed when the evolved data are fitted using  the Hubble diagram or
the angualr diameter test are due to the fact that, given the observed
magnitudes   and apparent angles, an    increase with redshift in  the
standard candle absolute luminosity  causes the best fitting distances
to be biased towards higher values, while  a decrement in the physical
size implies that real cosmological distances are underestimated.

Moreover,  even  considering  a  linear evolutionary  picture  for the
absolute luminosity as well as a modest  change in the standard candle
luminosity,  i.e. $\Delta_M=M(z)-M(0)=-0.5$    at   z=1.5, the   input
fiducial cosmology  falls  outside  the  3$\sigma$ confidence  contour
obtained  by applying the   magnitude-redshift test  to the  sample of
evolved   standard candles. Note that error   contours  are derived by
assuming a scatter of $\sigma_{M}=0.05$ per redshift bin in the Hubble
diagram.  Since, at  variance with the size  of large  discs, galactic
luminosity is expected to evolve  substantially with redshift  (within
the VVDS survey, $\Delta_M  \sim -1$ in the  I band for M$^*$ galaxies
\citep{ilbe05}) we conclude  that  the  direct implementation  of  the
Hubble diagram   test  as a  minimial test  for   the parameter subset
[$\Omega_Q,w$] using galaxy  rotation as the standard candle indicator
is more problematic.  As an additional problem, we note that
 galaxy luminosity is seriously  affected by uncertainties in internal
 absorption
corrections,  and that due to K-correction,  the implementation of the
Hubble  diagram requires  multi  band images  to properly describe the
rest frame emission properties of galaxies.

\subsection{The cosmology-evolution diagram}

In this  section  we  want  to address the  more   general question of
whether it is possible to infer  cosmological information knowing {\it
a-priori} only  the  upper limit   value for  disc   evolution at some
reference  redshift (for   example, the maximum  redshift
surveyed by a  given sample of  rotators).  In other terms, we explore
the  possibility of  probing geometrically  the cosmological parameter
space in a way which is independent of the specific evolution function
with which disc sizes change as a function  of time. The only external
prior is the knowledge of  an upper limit  for  the amplitude of  disc
evolution at some past epoch.

Given an arbitrary model specified by a set of cosmological parameters
$\vec{p}$,  and   given  the observable   $\theta^{obs}(z)$,  i.e. the
apparent  angle subtended by   a sample of velocity selected  galaxies
(with locally calibrated diameters $D_v(0)$), then

\begin{equation}         \epsilon_{\theta}(z,\vec{p})=\theta^{obs}(z)-
\theta^{th}(D_v(0),z,\vec{p}) \label{2} \end{equation}

is the function which describes the redshift evolution of the standard
rod, i.e     $\epsilon_{\theta}=(D_v(z)-D_v(0))/d_{\theta}$    in  the
selected cosmology.

Let's suppose that we know the lower and upper  limits of the relative
(adimensional) standard        rod          evolution    
($\delta (\bar{z})=\frac{|\Delta D_v(\bar{z})|}{D_v(0)}$) at some specific 
redshift $\bar{z}$.

Assuming  this prior, we    can  solve for  the  set  of  cosmological
parameters (i.e. points $\vec{p}$ of the cosmological parameter space)
which satisfy the condition

\begin{equation}   \delta_{l}(\bar{z})  \leq  \frac{|\epsilon_{\theta}
(\vec{p},        \bar{z})|}{\theta^{th}(D(0),\bar{z},\vec{p})}    \leq
\delta_u(\bar{z}). \label{cond4} \end{equation}

This inequality establishes a mapping between cosmology and the amount
of disc evolution  at  a given  redshift which is  compatible with the
observed  data.  By  solving it, one   can construct a self-consistent
cosmology-evolution  plane  where to   any given  range  of  disc size
evolution at $\bar{z}$ corresponds  in a unique  way a specific region
of the cosmological parameter space.  {\it Vice  versa}, for any given
cosmology one can extract  information about disc evolution.  Clearly,
the scatter in the angular-diameter diagram directly translates in the
uncertainties    associated   to  the  evolution     boundaries in the
cosmological space.

We note  that   the boundaries  of   the  region of   the cosmological
parameter space which  is compatible  with the  assumed prior  on  the
evolution  of diameters at $\bar{z}$ can  be equivalently expressed in
term of the maximum absolute evolution in luminosity (see appendix A).
This because, as  stated  in the  previous  section, velocity selected
objects have the unique property  of being at  the same time standards
of reference both in size and luminosity.

With this approach, one may by-pass the lack of knowledge about of the
particular evolutionary track  of  disc scalelengths and  luminosities
and try to extract information  about cosmology/evolution by giving as
a prior  only  the fractional evolution  in  diameters or the absolute
evolution in  magnitude expected at a  given redshift.  The essence of
the method is as  follows: instead of  directly putting constraints in
the cosmological  parameters space by  mean of cosmological probes, we
study how  bounded  regions in  the evolutionary plane  ($\frac{\Delta
D}{D}, \Delta M$) map onto the cosmological parameter space.

\begin{figure} \includegraphics[width=94mm,angle=0]{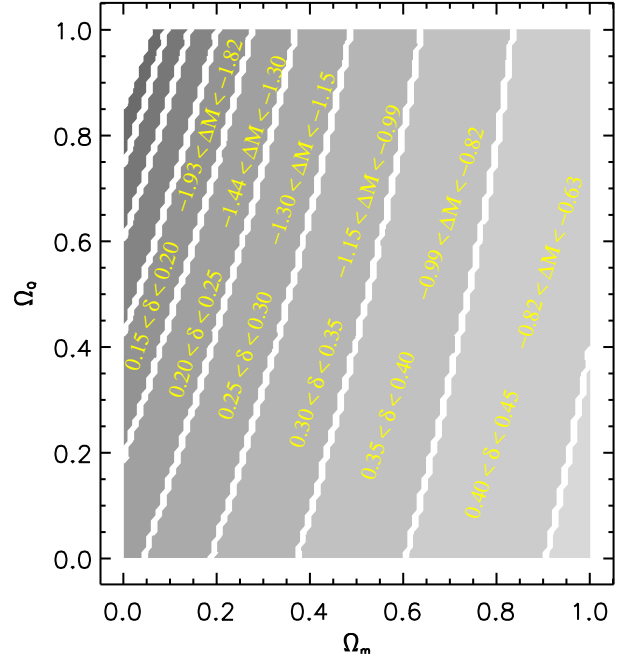} \caption{
Cosmology-evolution diagram for  simulated data which are affected  by
evolution.  Apparent angles and luminosities  of the velocity selected
sample of rotators are simulated in a $\Lambda$CDM cosmology. Standard
rods  and  candles have  been    artificially   evolved  so  that   at
$\bar{z}=1.5$ discs   are   26$\%$ smaller and  luminosities   1.4 mag
brigther.  The cosmological     plane is  partitioned  with  different
boundaries obtained  by  solving  equation \ref{cond4}  for  different
values of $\delta(\bar{z}=1.5)$, i.e. of the external prior representing
the guessed upper limit of the relative  disc evolution at the maximum
redshift covered by data. The external prior is also expressed in term
of absolute luminosity    evolution (see discussion in   Appendix A).}
\label{fig9} \end{figure}

In Fig \ref{fig9} we show  the cosmology-evolution diagram derived  by
solving eq. \ref{cond4} for  different ranges of $\delta(\bar{z})$. The
reference  model    is    the  concordanace   model   ($\Omega_m=0.3$,
$\Omega_{Q}=0.7$, $w=-1$) and the reference evolution at $\bar{z}=1.5$
is assumed  to be $\delta(\bar{z})=0.25$ for
discs,     and  $M_{v}(\bar{z})-M_{v}(0)$=-1.5 for  luminosities.  The
cosmology-evolution  diagram   represents   the unique  correspondance
between   all  the possible  cosmological models   and   the amount of
evolution in size and luminosity which is compatible with the observed
data at the given  reference redshift. Using independent  informations
about the range of evolution  expected in the structural parameters of
galaxies, for exemple from  simulations or theoretical models, one may
constrain the value  of cosmological parameters  {\it Vice versa},  if
the  cosmological model is known, then  one may directly determine the
evolution in magnitude and  size  of the  velocity selected sample  of
rotators.

With  this  approach,  the   possibility  of   discriminating  between
different cosmologies depends on the amount of evolution affecting the
standard rods at redshift  $\bar{z}$. Two different  sets ($\vec{p_1}$
and $\vec{p_2}$)  of cosmological parameters  may  be discriminated at
$z=\bar{z}$ if the relative disc evolution at $\bar{z}$  is known to a
precision better than

\begin{equation}    \delta(\bar{z})<       \frac{|r(\bar{z},\vec{p_2})-
r(\bar{z},\vec{p_1})|}{r(\bar{z},\vec{p_1})} \end{equation}

For example, this kind of  analysis  shows that an Einstein-de  Sitter
universe  may be  unambigously discriminated from  a critical universe
with  parameter  $\Omega_m=0.3, \Omega_Q=0.7,  w=-1.$  if the relative
disc evolution at $\bar{z}=1.5$ is known  with a precision better than
$28\%$.  In a   similar   way, an open   ($\Omega_m=0.3,  \Omega_Q=0$)
universe  may be discriminated from an  Einstein de-Sitter universe if
$\delta(\bar{z}) < 17\%$.

\begin{figure} \includegraphics[width=84mm,angle=0]{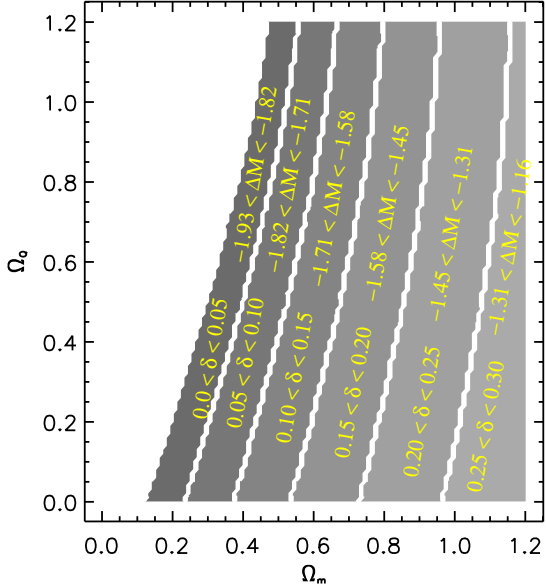} \caption{
As in fig. \ref{fig9}  but for a  different cosmology (an  Einstein-de
Sitter universe).}  \label{fig10} \end{figure}

\section{Conclusions}

The scaling of the apparent angular diameter of galaxies with redshift
$\theta(z)$ is  a powerful discriminator  of  cosmological models. The
goal of  this  paper  is  to    explore the   potentiality of a    new
observational  implementation  of the  classical angular-diameter test
and to study its performances and limitations.

We propose to use the velocity-diameter relationship, calibrated using
the [OII]$\lambda 3727$\AA~ line-widths,  as a tool to select standard
rods and probe world  models. As for  other purely geometrical test of
cosmology,   a  fair  sampling  of   the   galaxy  population  is  not
required.  It is however imperative  to have high quality measurements
of the structural parameters of high redshift galaxies (disc sizes and
rotational   velocity).   Surveys with HST    imaging  and high enough
spectral resolution will thus  provide the fundamental ingredients for
the practical realization of the recipe we have presented.

In  order to avoid any luminosity  dependent selection effect (such as
for exemple Malmquist bias) it is necessary to apply the proposed test
to  high velocity rotators.   We show  that nearly  1300 standard rods
with rotational velocity  in the bin $V  \sim 200\pm  20$ km s$^{-1}$)
are  expected in a field  of size  1 deg$^2$  over  the redshift range
$0<z<1.4$. Interestingly this large sample can be quickly assembled by
the currently underway zCOSMOS   deep redshift survey, which  uses the
VIMOS   multi-object spectrograph  at   the  VLT  to   target galaxies
photometrically selected using high-resolution ACS images.

Even allowing a  scatter   of $40\%$ for  the  [OII]$\lambda 3727$\AA~
linewidth-diameter  relationship for disc galaxies,   we show that the
angular-diameter diagram constructed using this  sample is affected by
a scatter  of  only $ \sim  5\%$ per  redshift  bin of amplitude
$dz=0.1$. This   scatter translates into  a   $20\%$ precision in  the
"geometric" measurement of the dark energy  constant equation of state
parameter $w$, through a test  performed without  {\it priors} in  the
[$\Omega_{m}, \Omega_{Q}$] space.

Current  theoretical  models suggest   that  large  discs (i.e.   fast
rotators)  evolve  weakly with  cosmic time  from  $z=1.5$ down to the
present epoch. Anyway, we  have explored how  an eventual evolution of
the velocity-selected standard rods might affect the implementation of
the test.   We have shown that any  possible evolution in the standard
rods may be   unanbiguously revealed by  the  fact that even   a small
decrement with redshift of the disc sizes shifts the inferred value of
the matter density parameter   into {"\em a-priori}  excluded" regions
($\Omega_m<0.2$).

We  have shown that  a linear  (as expected on  the  basis  of various
theoretical models)  and   substantial (up to   $40\%$  over the range
$0<z<1.5$)   disc evolution minimally biases   the inferred value of a
dark energy    component that  behaves  like   Einstein's cosmological
constant $\Lambda$. Moreover  we have shown  that assuming that  discs
evolve in a linear-like way as a function  of redshift, and that their
sizes were  not more than  $30\%$ smaller  at $z=1.5$ with  respect to
their present  epoch dimension, then  the angular diameter test can be
used  to  place  interesting  constraints  in  the  [$\Omega_{Q},  w$]
plane. In particular, assuming a scatter  of $5\%$ per redshift bin in
the  angular  diameter-redshift  diagram (nearly  corresponding to the
scatter expected  for   a sample  of   1300  rotators with  $0<z<1.4$,
$dz=0.1$, whose diameter   is    locally calibrated with  a     $40\%$
precision), we have shown that the input fiducial $[\Omega_Q,w]$ point
is still within the 1$\sigma$ error contours  obtained by applying the
angular diameter test to the evolved data.

Finally, we have outlined the strategy to derive a cosmology-evolution
diagram with which it is possible to  establish an interesting mapping
between different cosmological   models  and  the amount  of    galaxy
disc/luminosity evolution     expected  at  a  given    redshift.  The
construction of this  diagram   does not  require an  {\it   a-priori}
knowledge of  the     particular  functional  form  of   the    galaxy
size/luminosity evolution. By   reading  this diagram, one   can infer
cosmological  information  once  a    theoretical  prior  on   disc or
luminosity evolution at a given redshift  is assumed. In particular if
the amplitude of the relative disc evolution at $\bar{z}=1.5$ is known
to  better than   $\sim 30\%$, then   an  Einstein-de Sitter  universe
($\Omega_m=1$) may be geometrically  discriminated from a flat, vacuum
dominated one ($\Omega_m=0.3$,  $\Omega_Q=0.7$).  {\it Viceversa}, one
can use  the cosmology-evolution diagram to  place constraints  on the
amplitude of the  galaxy disc/luminosity  evolution, once a  preferred
cosmology is chosen.

In conclusion, given    the simple ingredients entering   the proposed
implementation strategy, nothing, besides evolution of discs, could in
principle bias  the test. Even so,  evolution can be  easily diagnosed
and, under some general conditions,  it can be  shown that it does not
compromise the  possibility of detecting the  presence of  dark energy
and constraining the value of its equation of state.

In the following papers  of this series \citep[]{paperII,paperIII}, we
implement the    proposed  strategy   to  a  preliminary    sample  of
velocity-selected high redshift rotators.

\section*{Acknowledgments}  We   would   like to   acknowledge  useful
discussions with R. Scaramella and G. Zamorani.
This  work has been partially  supported by NSF grants AST-0307661 and
AST-0307396 and was done while AS was receiving  a fellowship from the
{\it   Fonds  de  recherche sur la    Nature  et les   Technologies du
Qu\'{e}bec}.  KLM is supported by the NSF grant AST-0406906.

\section{Appendix A}

Given a spectroscopically selected   sample of objects   with constant
rotational velocity we can derive the observed  magnitude $m^{o}$ of a
standard candle of absolute magnitude $M_v(0)$  located at redshift z,
by     using the    standard    relation \cite{SAN}   \begin{equation}
m^{o}=m^{th}(M_v(0),z,\vec{p})+\epsilon_M(z)+K(z)            \label{1}
\end{equation}

where \[m^{th}=M_v(0)+5\log d_{L}(z,\vec{p})+25\]  and were $d_{L}$  is
the luminosity distance, K(z) is the K correction term and
  $\epsilon_M(z)$ is the {\em a-priori} unknown
evolution      in    luminosity    of      our   standard      candle,
i.e. $\epsilon_M(z)=\Delta  M_v(z)=M_v(z)-M_v(0)$ is   the  difference
between the absolute  magnitude of an object  of rotational velocity V
measured at redshift $z$    and the un-evolved local standard    value
$M_v(0)$.  

From the definition of wavelength-specific surface brightness $\mu$ we
deduce that the variation  as a  function of  redshift in  the average
intrinsic  surface  brightness  (within a   radius R)  for our  set of
homologous galaxies is

\[  \Delta   \langle\mu^{th}(z)\rangle_R=\Delta M_v(<R)      -5   \log
\frac{R(z)}{R(0)} \]

By opportunely  choosing  the  half  light radius  $D_v$  as a  metric
definition for the size of a galaxy we immediately obtain

\begin{equation}   \Delta \langle\mu^{th}(z)\rangle_D=\epsilon_M(z,\vec{p})
-5                                                                \log
\Big(\frac{\epsilon_{\theta}(z,\vec{p})}{\theta^{th}(D(0),z,\vec{p})}+1\Big)
\label{rel} \end{equation}

The  intrinsic surface brightness evolution is  not an observable, but
in a FRW  metric this  quantity is  related to the  surface brightness
change observed in a waveband $\Delta \lambda$ by the relation

\begin{equation} \Delta \langle\mu^{o}(z)\rangle_D=\Delta \langle
\mu^{th}(z)\rangle_D +2.5 \log(1+z)^4+K(z) \label{4} \end{equation}

Thus,  once   we  measure   the     redshift evolution  of     $\Delta
\langle\mu^{o}(z)\rangle_D$ for  the sample of rotators,  the absolute
evolution in luminosity corresponding to a given relative evolution in
diameters can be directly inferred using eq. \ref{rel}.

\label{lastpage}


\begin{thebibliography}{99}

\bibitem[\protect\citeauthoryear{Aldering     et    al.}{2002}]{ald02}
Aldering,     G.  et  al. 2002,  in      SPIE Proceedings   Vol 4835,
astro-ph/0209550


\bibitem[\protect\citeauthoryear{Berlind   \&  Weinberg}{2002}]{ber02}
Berlind, A. A., \& Weinberg, D. H. 2002, \apj, 575, 587

\bibitem[\protect\citeauthoryear{Boissier  \&  Prantzos}{2001}]{boi01}
Boissier, S., Prantzos, N. 2001, MNRAS, 325, 321

\bibitem[\protect\citeauthoryear{Bottinelli    et   al.}{1980}]{bot80}
Bottinelli,  L., Gouguenheim,   L., Paturel, G.,  \& de Vaucouleurs,  G.
1980, ApJ, 242, L153

\bibitem[\protect\citeauthoryear{Bouwens     \&    Silk}{2002}]{bou00}
Bouwens, R., \& Silk, J. 2000, ApJ, 568, 522


\bibitem[\protect\citeauthoryear{Bruzual    \&  Spinrad}{1978}]{bru78}
Bruzual, A. G., \& Spinrad, H. 1978, ApJ, 220, 1

\bibitem[\protect\citeauthoryear{Buchalter   et     al.}{1998}]{buc98}
Buchalte, A., Helfand, D. J., Becker, R. H., \& White,  R. L. 1998, ApJ,
494, 503

\bibitem[\protect\citeauthoryear{Carroll,              Press        \&
Turner}{1992}]{car92} Carroll,  S.  M.,  Press,   W. H.,  \&   Turner,
E. L. 1992, ARA\&A, 30, 499


\bibitem[\protect\citeauthoryear{Chiappini,        Matteucci        \&
Gratton}{1997}]{chia97}   Chiappini,  C.,    Matteucci,   F., \& Gratton,
R. 1997, ApJ, 477, 765


\bibitem[\protect\citeauthoryear{Colless      et    al.}{1998}]{col98}
Colless, M., Glazebrook, K., Mallen-Ornelas, G., \& Broadhurst, T.  1998,
ApJ, 500, L75


\bibitem[\protect\citeauthoryear{Cooray et  al.}{2001}]{coo01} Cooray,
A., Hu, W., Huterer, D., \& Joffre, M.  2001, ApJ, 557L, 7


\bibitem[\protect\citeauthoryear{Daly}{2004}]{da04} Daly, R. A.  2004,
\apj, 612, 652


\bibitem[\protect\citeauthoryear{Daly    \&   Djorgovski}{2004}]{dadj}
Daly, R. A., \&  Djorgovski, S. G.  2004, \apj, 612, 652


\bibitem[\protect\citeauthoryear{Davis et al.   }{2000}]{dav00} Davis,
M., Newman, J. A., Faber, S. M., \& Phillips, A. C.  2000 in Deep Fields,
Proceedings   of the   ESO/ECF/STScI Workshop,   eds.   Cristiani, S.,
Renzini, A., Williams, R. E., Springer, p. 241




\bibitem[\protect\citeauthoryear{de Bernardis et al.}{2002}]{deb02} de
Bernardis, P.  et al.  2002, ApJ, 564, 559

\bibitem[\protect\citeauthoryear{de  Vaucouleurs,  de  Vaucouleurs, \&
Corwin}{1976}]{dev76} de Vaucouleurs,  G., de Vaucouleurs, G., \& Corwin,
H.   G.  1976,     {\it  Second  Reference   Cataloguwe      of Bright
Galaxies}(Austin: University of Texas) RC2


\bibitem[\protect\citeauthoryear{Djorgovski  \& Spinrad}{1981}]{djo81}
Djorgovski, S., \& Spinrad, H.  1981, ApJ, 251, 417

\bibitem[\protect\citeauthoryear{Faber   et  al.}{2001}]{fab01} Faber,
S. M.,  Phillips, A. C., Simard, L.,  Vogt, N. P., \&  Somerville, R. S.,
2001,   in  Galaxy Disks  and  Disk Galaxies,   ASP Conference Series,
Vol. 230, p. 517

\bibitem[\protect\citeauthoryear{Ferguson   \&   Clarke}{2001}]{fer01}
Ferguson, A. M. N., \& Clarke, C. J.  2001, MNRAS, 325, 781


\bibitem[\protect\citeauthoryear{Gonzales}{2002}]{gon02}     Gonzalez,
A. H.  2002, ApJ, 567, 144


\bibitem[\protect\citeauthoryear{Gurvits,        Kellermann         \&
Frey}{1999}]{gur99} Gurvits, L. I., Kellermann, K. I., \& Frey, S.  1999,
A\&A, 342, 378

\bibitem[\protect\citeauthoryear{Haiman,        Mohr                \&
Holder}{2001}]{haim01}, Haiman,   Z.,   Mohr,  J.   J.,   \&   Holder,
G. P. 2001, ApJ, 553, 545

\bibitem[\protect\citeauthoryear{Halverson    et    al.}{2002}]{hal02}
Halverson, N. W., et al. 2001, ApJ, 568, 38

\bibitem[\protect\citeauthoryear{Hickson}{1977}]{hic77}  Hickson,  P.,
1977a, ApJ, 217, 16

\bibitem[\protect\citeauthoryear{Hogg}{1999}]{hog99}  Hogg, D. W.
1999, astro-ph/9905116

\bibitem[\protect\citeauthoryear{Hoyle}{1959}]{hoy59} Hoyle, F.,  1959
in Proc. IAU Symp. 9 and URSI Symp. 1,  Paris Symp on Radio Astronomy,
ed. R. N. Brachewell (Stanford: Stanford Univ. Press), 529

\bibitem[\protect\citeauthoryear{Hubble \&  Tolman}{1935}]{HT} Hubble,
E., \& Tolman, R. C., Astrophys. J., 1935, 82, 302
 
\bibitem[\protect\citeauthoryear{Huterer    \&   Turner}{2000}]{hut00}
Huterer, D., \& Turner, M. S.  2001, Phys.Rev. D64

\bibitem[\protect\citeauthoryear{Ilbert et al.}{2005}]{ilbe05} Ilbert,
O., et al. 2005, A\&A, 439, 863

\bibitem[\protect\citeauthoryear{Kapahi}{1975}]{kap75}  Kapahi, V. K.,
1975, MNRAS, 172, 513

\bibitem[\protect\citeauthoryear{Kellermann}{1993}]{kel93} Kellermann,
K. I.  1993, nature 361, 134

\bibitem[\protect\citeauthoryear{Kobulnicky \& Gebhardt}{1999}]{kob99}
Kobulnicky, H. A., \& Gebhardt, K., ApJ, 1999, 119, 1608


\bibitem[\protect\citeauthoryear{Lahav et  al.}{2002}]{lah02}   Lahav,
O.  2002, \mnras, 333, 961

\bibitem[\protect\citeauthoryear{Lee  et al.}{2001}]{lee01} Lee, A. T.
et al. 2001, ApJ, 561, L1

\bibitem[\protect\citeauthoryear{Le F\`evre et al.}{2005}]{lef05}   Le
F\`evre, 0. et al.  2005, A\&A, 439, 845

\bibitem[\protect\citeauthoryear{Lilly   et  al.}{1998}]{lil98} Lilly,
S. J., Schade, D.,  Ellis, R., Le Fevre,  O., Brinchmann,  J., Tresse,
L.,  Abraham, R., Hammer,  F.,  Crampton, D., Colless, M., Glazebrook,
K., Mallen-Ornelas, G., \& Broadhurst, T.  1998, ApJ, 500, L75

\bibitem[\protect\citeauthoryear{Lilly et al.}{2006}]{lil06}    Lilly,
S. J., et al. 2006, \apj, in press (astro-ph/0612291)

\bibitem[\protect\citeauthoryear{Lima \&  Alcaniz}{2002}]{lim02} Lima,
J. A. S., \& Alcaniz, J. S.  2002, A\&A, 566, 15L

\bibitem[\protect\citeauthoryear{Marinoni    et     al.}{1998}]{mar98}
Marinoni,  C., Monaco, P.,  Giuricin,  G., \& Costantini, B.  1998, \apj,
505, 484

\bibitem[\protect\citeauthoryear{Marinoni    \&  Hudson}{2002}]{mar02}
Marinoni, C. \& Hudson, M.  2002, \apj, 569, 101

\bibitem[\protect\citeauthoryear{Marinoni     et    al.}{2005}]{mar05}
Marinoni, C., et al. 2006, A\&A, 442, 801

\bibitem[\protect\citeauthoryear{Marinoni   et   al.}{2007}]{paperIII}
Marinoni, C. et al. 2007, A\&A in press, astro-ph/0710.0761

\bibitem[\protect\citeauthoryear{Miley}{1971}]{mil71}    Miley, G. K.,
1971, MNRAS, 152, 477

\bibitem[\protect\citeauthoryear{Mo,   Mao    \& White}{1998}]{MO} Mo,
H. J., Mao, S., \& White, S. D. M.  1998, MNRAS, 295 319


\bibitem[\protect\citeauthoryear{Narayan    \&    White}{1988}]{nar88}
Narayan, R., \& White, S. D. M.  1988, MNRAS, 231, 97

\bibitem[\protect\citeauthoryear{Newman et  al.}{2002}]{new02} Newman,
J, A., Marinoni, C., Coil, A. L., \& Davis, M.  2002, PASP, 114, 29

\bibitem[\protect\citeauthoryear{Newman      \&   Davis}{2000}]{new00}
Newman, J. A., \& Davis, M.  2000, ApJ, 534, L11

\bibitem[\protect\citeauthoryear{Nelson et  al.}{2002}]{nel02} Nelson,
A, E.,  Simard, L., Zaritsky, D., Dalcanton,  J. J.,  \& Gonzalez, A. H.,
2002, ApJ, 567, 144

\bibitem[\protect\citeauthoryear{Pen}{1997}]{pen97}  Pen,   U-L. 1997,
New A., 2, 309

\bibitem[\protect\citeauthoryear{Perlmutter    et   al.}{1999}]{per99}
Perlmutter, S. et al. 1999, ApJ, 517, 565

\bibitem[\protect\citeauthoryear{Petrosian}{1976}]{pet76}   Petrosian,
V.  1976, ApJ, 209, L1

\bibitem[\protect\citeauthoryear{Riess  et   al.}{2001}]{rie01} Riess,
A. G. et al. 1999, ApJ, 560, 49

\bibitem[\protect\citeauthoryear{Saintonge   et   al.}{2007}]{paperII}
Saintonge, A., Masters, K.L., Marinoni, C., Giovanelli, R., \& Haynes,
M.P. 2007, A\&A submitted, astro-ph/0710.0760

\bibitem[\protect\citeauthoryear{Sandage}{1972}]{san72}   Sandage, A.,
ApJ,  1972, 173, 485

\bibitem[\protect\citeauthoryear{Sandage}{1972}]{SAN}  Sandage,    A.,
1998, Ann. Rev. Astron. Astrophys. 26, 561

\bibitem[\protect\citeauthoryear{Sandage}{1995}]{san95}  Sandage,  A.,
1995, in  Saas-Fee Advanced Course   23, The Deep Universe:  Practical
Cosmology; ed. B. Binggeli \& R.  Buser (New York: Springer)

\bibitem[\protect\citeauthoryear{Spergel     et     al.}{2006}]{spe06}
Spergel, D. N., et al.  2006, astro-ph/0603449

\bibitem[\protect\citeauthoryear{Springob     et     al.}{2007}]{spri07}
Springob,  C.~M.,  Masters, K.~L., Haynes,  M.~P.,  Giovanelli, R., \&
Marinoni, C. 2007,  ApJS in press,  astro-ph/07050647

\bibitem[\protect\citeauthoryear{Takamiya}{1999}]{tak99} Takamiya, M.,
1999 ApJS, 122, 109

\bibitem[\protect\citeauthoryear{Tegmark}{2006}]{teg06} Tegmark, M.
2006, Phys. Rev. D 74, 123507 

\bibitem[\protect\citeauthoryear{Tolman}{1930}]{tol30} Tolman,  R. C.,
1930, Proc. Natl. Acad. Sci. USA: 16, 511

\bibitem[\protect\citeauthoryear{Totani  et al.}{2002}]{tot02} Totani,
T., Yoshii,  Y., Maihara,  T., Iwamuro,  F., \& Motohara,  K.  2001, ApJ,
559, 592

\bibitem[\protect\citeauthoryear{Tully \& Fisher}{1977}]{tul77} Tully,
R. B., \& Fisher, J. R.  1977, A\&A, 54, 661

\bibitem[\protect\citeauthoryear{Turner      \&   White}{1997}]{tur97}
Turner, M. S., \& White, M.  1997, Phys. Rev. D, 56, 4439

\bibitem[\protect\citeauthoryear{van  den Bosch}{2006}]{vdb06} van den
Bosch, F. C., et al. 2006, MNRAS submitted (astro-ph 0610686)

\bibitem[\protect\citeauthoryear{Wilkinson     et   al.}{1998}]{wil98}
Wilkinson,  P.N.,  Browne,  I.  W. A.,   Alcock,  D.  et al.  1998  in
Observational   Cosmology    with   new      radio   surveys,  Kluwer,
Acad. Publishers, Dordrecht, p. 221
 
\end{thebibliography}
\end{document}